\documentclass[aps,nofootinbib,onecolumn,notitlepage,11pt]{revtex4-2}

\usepackage{amssymb, amsmath, amsthm, amsfonts}
\usepackage{graphicx}
\usepackage{color}
\usepackage{mathrsfs}
\usepackage{physics}
\usepackage{mathtools}

\usepackage[UKenglish]{isodate}
\usepackage[UKenglish]{babel}

\usepackage{hyperref}

\begin{document}

\title{Formalising the Use of the Activation Function in Neural Inference}

\author{Dalton A R Sakthivadivel}
\email{dalton.sakthivadivel@stonybrook.edu}

\affiliation{VERSES Research Lab, Los Angeles, CA, 90016, USA}

\affiliation{Departments of Mathematics, Physics and Astronomy, and Biomedical Engineering, Stony Brook University, Stony Brook, NY, 11794-3651, USA}

\date{\today}

%%%%%%%%%%%%%%%%%%%%%%%%%%%%%%%%%%%%%%%%%%%%%%%%%%%%%%%%%%%%%%%%%%%%%%%

\begin{abstract}

We investigate how the activation function can be used to describe neural firing in an abstract way, and in turn, why it works well in artificial neural networks. We discuss how a spike in a biological neurone belongs to a particular universality class of phase transitions in statistical physics. We then show that the artificial neurone is, mathematically, a mean field model of biological neural membrane dynamics, which arises from modelling spiking as a phase transition. This allows us to treat selective neural firing in an abstract way, and formalise the role of the activation function in perceptron learning. The resultant statistical physical model allows us to recover the expressions for some known activation functions as various special cases. Along with deriving this model and specifying the analogous neural case, we analyse the phase transition to understand the physics of neural network learning. Together, it is shown that there is not only a biological meaning, but a physical justification, for the emergence and performance of typical activation functions; implications for neural learning and inference are also discussed.

\end{abstract}

%%%%%%%%%%%%%%%%%%%%%%%%%%%%%%%%%%%%%%%%%%%%%%%%%%%%%%%%%%%%%%%%%%%%%%%

\maketitle

\section{Introduction}

%What do we know -- lit review
The perceptron learning algorithm, developed by McCulloch and Pitts in 1943, is one of the earliest applications of biological principles for computation to mathematics, or to machines \cite{Abraham}. A simple model, the perceptron consists of a single logic gate, and is only capable of classification using linearly separable functions, like {\large\scshape and} and {\large\scshape or}. Nonetheless, recent algorithms have deviated only slightly from the original developments by McCulloch and Pitts; in many cases, these simply stack perceptrons or add features onto the original algorithm, such as in deep neural networks or convolutional neural networks. Clearly, the contribution of the single-layer perceptron remains relevant today.

%what we don't know -- questions, problems, or gaps
Somewhat anomalous in the perceptron, and indeed in further models, is the critical importance of the activation function. McCulloch and Pitts recognised that neural firing occurs in an all-or-none fashion, and that any function with a rapid transition between two end behaviours would suffice to describe this phenomenon \cite{MP1943}. The same argument was later presented at the level of neural populations in \cite{WilsonCowan1972}, by showing that for a realistic distribution of single neurone activation, the ensemble activity necessarily looks like a sigmoid function. In other words, a specific class of functions is generally used for an activation function, which can be described as discontinuous or nearly discontinuous at a `switching point,' vertically asymmetric about this point, and bounded from below. Concrete examples include the Heaviside function originally used by McCulloch and Pitts, and Wilson and Cowan's sigmoid function. Interestingly, a class of activation functions that are bounded from below but exhibit asymptotically linear behaviour for inputs greater than a critical threshold, such as ReLU, ELU, Mish, and Swish, have been experimentally evaluated as providing the best performance for a large number of network architectures and tests \cite{NeurIPS2012,ICML,pmlr}. Much like the Heaviside activation function, however, these functions are justified by their performance, and are given heuristically.

Whilst the use of an activation function is and has been justified by the biological facts, and its success is obvious, it is still assembled primarily phenomenologically. The activation function was certainly integral to the application of neural networks as logical devices that classify non-binary variables\textemdash but, the precise mechanism that justifies these functions' role in inference, and the physiological relevance of this function, both remain unclear. Most proofs of the previous statement also yield little insight into the relevance of the activation function, and especially of the specific shape elaborated on above. These proofs often rely on what could be summarised as the power of non-linearity, which allows the approximation of non-linear or non-polynomial functions. Consider that data-generating processes are governed by a dynamical system, which could be a high-dimensional stochastic system or partial differential equation, the solutions to which are typically non-linear or non-polynomial in character. Then, the necessity of such a function becomes clear. In greater detail, a theorem offered in \cite{pinkus1993} states that the set of possible neural network configurations $\mathcal{N}$ is \emph{dense} in the space of continuous real-valued functions, or, that any real-valued function is contained in or is a limit point of $\mathcal{N}$, if and only if the activation function on $\mathcal{N}$ is non-polynomial. In other words, given arbitrary width and depth, the property of being a `universal approximator' is precisely that of having a non-polynomial activation function. Still, this proof leads to little insight about the biological plausibility of, or physical motivation for, the specific functions used.

%what we're doing about it -- how we solve it (methods)
To understand how activation functions arise in artificial neural networks, and how they are connected to the fundamentals of biologically-inspired computation, we employ a model from statistical mechanics called the Ising model. The Ising model was devised by Wilhelm Lenz and Ernst Ising in the 1920s to describe magnetism in metals, and the loss of magnetisation when magnets are heated \cite{Ising67}. The Ising model is a model of the atomic structure of a metal, where the nuclei of metallic atoms are defined with a property called `spin,' pointed either up or down. When all spins are positively aligned, or, all lattice sites take values of $+1$, the system is magnetised. Like many systems in nature, it exhibits a phase transition at a critical temperature, in which the magnetisation of a cooled metal is lost above a critical temperature, or regained when cooled. In fact, many other phase transitions can be proven to take characteristics of this one\textemdash phase transitions lie in one of a few \emph{universality classes}, meaning they show the same characteristics no matter what the underlying dynamics are. The Ising universality class contains such different phenomena as the liquid-gas critical point, and the behaviour of strings in string theory \cite{liq-gas-crit,IsingStrings}. Similar to this class of behaviour, a neural spike is a sort of transition where, once past a critical point (a threshold potential) a spike is initiated, and the system goes from disordered uptake (random diffusion of ions across the membrane) to ordered uptake (uptake of $\text{Na}^+$ and other positive ions to initiate depolarisation). We will use this correspondence to model the spiking behaviour of the neural cell as an Ising model with an appropriate phase transition. In so doing, we recover expressions for both the hyperbolic tangent and unbounded-above linear activation function. This shows that an artificial neural network must be equipped with such an activation function if it is to be a meaningful approximation of a biological neural network, offering a new take on the manner in which activation functions lead to neural computation.

\section{Main Results}

\subsection{Modelling Neural Dynamics}\label{TwoA}

Neural spikes are both a regular phenomenon, and a highly complex, non-equilibrium process. As an example of self-organisation, neural firing emerges from complex but quantifiable dynamics, here involving ionic equilibria and membrane selectivity \cite{HH1952}. The neurone is surrounded by ions in its extracellular fluid, meaning it is subject to diffusion of these ions across its cell membrane, through ion channels. It maintains a negative resting potential of around $-70$ mV, which requires active transport of positive $\text{Na}^+$ ions out of the cell. This results in a persistent concentration gradient, which is precisely what allows a spike to occur. When a critical voltage is reached, previously closed voltage-gated ion channels open. The positive $\text{Na}^+$ ions flow along this concentration gradient through these now open channels, leading to an upwards spike in voltage.

We have simplified this model of neural dynamics to be a stationary process coupled to a bath. We can then approximate the system as being in a local equilibrium, meaning we can use a simple equilibrium Ising model to describe the system. Consider a particular formulation of the Ising model coupled to a thermal bath, which undergoes a rapid quench and magnetises in response to this sudden cooling. When the quench is removed, the Ising model heats up again. For periodic quenching, the dynamics themselves will be periodic, but will obey the typical $h=0$ transition in magnetisation. The phases induced by the quench can thus be made distinct, {\it e.g.}, alternating disordered phases with thermal fluctuations and ordered phases with positive magnetisation. This provides a model of neural dynamics, in which the crucial simplification in the model is ignoring the source of the external quench, thereby restricting us only to local (intracellular) interactions within the system. There is no consequence to the validity of our model, since firing is the organisation of channel dynamics within the cell.

\subsection{The Ising Case for Neural Firing}\label{TwoB}

As stated, both systems are capable of exhibiting two, differently stereotyped dynamics, or `phases.' In the Ising model, one is a high temperature \emph{paramagnetic} phase, where the spins in the model are disorganised and unaligned, weakly correlated with one another, and subject to random fluctuations. The magnetisation $m$, or average spin, is zero in this case; this is a result of the random configuration of spins, such that approximately half of the spins should be occupying states $-1$ and half occupying $+1$, for $m = \langle s \rangle = 0$. 

We also observe a \emph{ferromagnetic} phase in which the spins are organised and aligned in one direction. Here, $m$ is either $-1$ or $+1$, which correspond to anti-ferromagnetism and ferromagnetism, respectively. A quench is a decrease in thermal energy, which causes the spins to align with each other so that the model occupies a low energy state and $m = \langle s \rangle = 1$. We will see that the energy in the Ising model depends on the interactions between spins. As such, to get the total energy, we take the negative sum of spin states over all pairs of neighbours. Clearly, when pairs of neighbouring spins are aligned in the positive direction, such that the sum of $n$ spin pairs is $-n$, the energy is at a minimum. The converse is also true: when the energy in the system decreases, spins will align and take a lower energy configuration to satisfy this. The final state, $-1$ or $+1$, is `chosen' as the magnet cools towards the critical temperature according to the boundary condition of the model. 

We consider the neural membrane a two-dimensional lattice of channels wrapped around a cell body. The transition from order to disorder, and subsequent ensemble action in spiking, is now that of the Ising model. It is well known that a threshold potential exists, in which a neurone exhibits sudden spiking in response to a critical level of stimulation. In this case, the state of the neurone\textemdash the voltage\textemdash is the negative of Ising temperature, and as voltage increases towards the threshold, temperature decreases towards a critical temperature $T_c$. In this case, the adaptation of the standard Ising model applies well. Whilst it is coupled to a thermal bath that increases its temperature, it remains disordered due to high energy fluctuations. Similarly, at rest, the neurone occupies a highly entropic and thus high energy state, given the open and closed channels in the membrane; it is also subject to fluctuations that destroy order in the system as a result of constant diffusion and pump dynamics. Together, the two maintain a disordered resting state, and along with the highly entropic channel states, a spike is impossible without cooling. 

We will treat events discretely, as a single spike in response to an input. We will also consider the neurone as being in disequilibrium with its surroundings, with occasional field interactions. As stated, for modelling purposes, we simplify this as a local input effect. In the neural case, cooling comes in the form of the summation of inputs from other neurones in the network. This plays the role of a quench, in that it moves the system's state towards order. 

\subsection{The Transition to Magnetisation}

To connect macroscopic observables to microscopic state variables, we often rely on formalisms from statistical mechanics. One such technique used in study of phase transitions is a particular type of coarse graining called mean field theory (MFT), which formulates a model of the macroscopic level change that results from certain microscopic changes. MFT is quantitatively incorrect in two-dimensions, and is only an approximation; nonetheless, for both computational and pedagogical reasons, we will demonstrate this using the mean field approach.

Here, we will briefly state the derivation for the phase transition in the Ising model. A spin lattice in zero field is described by its Hamiltonian $\hat H$ in the following way:
\[
\hat H = -J \sum_{i,j} s_i s_j,
\]
with $s_n \in \{ -1,1 \}$. $\hat H$ gives the total energy of the system, which in turn gives its dynamics, as the sum over all neighbouring spins. In our analogous neural model, a spin is a channel state, which at any time $t$ either contains an $\text{Na}^+$ ion or does not.

MFT assumes that at large length scales, a system converges to its average dynamics, with only large fluctuations playing a role in the dynamics of the system. We use this to coarse-grain the model by removing second order fluctuations, which are assumed to be vanishingly small. The local interaction term in $\hat H$ is 
\[
s_i s_j = (\langle s_i \rangle + \sigma s_i)(\langle s_j \rangle + \sigma s_j),
\]
which is a product of two variables under the influence of a random displacement. We assert that in this system, the spins tend towards a similar mean and fluctuations necessarily decrease; therefore, in describing the dynamics leading to an ordered transition, we may assume the fluctuations become small. This means that we can rewrite $s_i s_j$ as the following:
\[
\langle s_i \rangle \langle s_j \rangle + \langle s_j\rangle \sigma s_i + \langle s_i \rangle \sigma s_j + \sigma s_i \sigma s_j.
\]
Since we have assumed the fluctuations are small, the final term will vanish. If we use this, and an expansion of the random displacement into a fluctuation about a mean, this becomes
\[
s_i s_j \approx \langle s_i \rangle \langle s_j \rangle + \langle s_j\rangle (s_i - \langle s_i \rangle) + \langle s_i \rangle (s_j - \langle s_j \rangle).
\]
We also use the fact that as the phase transitions, the average spin value $\langle s_n \rangle$ will approach a magnetisation value $m$, corresponding to the organisation of spins needed to produce magnetisation. Then, we can rewrite $\hat H$ as
\[
-J \sum_{i,j} m(s_i + s_j) - m^2,
\]
replacing spins with the mean field $m$. If we take spin states as being highly correlated, then $i$'s and $j$'s become equal; in that case, the sum over neighbours will reduce to the number of connections, half the number of neighbours $z$, across all sites in the lattice. This gives a scaling factor of $\frac{z}{2}$:
\[
- \frac{zJ}{2} \sum_{i=1}^N 2ms_i - m^2.
\]
We can further simplify to
\[
\hat H_{\text{MF}} = - zJm\sum_{i=1}^N s_i  + \frac{NzJm^2}{2}
\]
by distributing the scaling factor into the sum.

\begin{figure}[h!]
\resizebox{0.45\textwidth}{!}{
\includegraphics{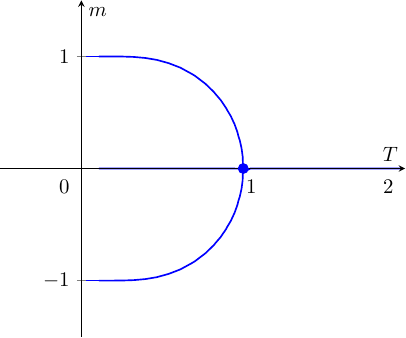}
}
\caption{\textbf{Mean field magnetisation $m$ as a function of temperature.} A phase transition is evident in the graph of \eqref{MFmodel} at $T=T_c$ (here set such that $T_c = 1$), where magnetisation becomes non-zero.}
\label{fig:m(T)}
\end{figure}

This mean field Hamiltonian describes what the total energy looks like at a phase transition, at a coarse scale. It does not, however, describe the transition itself. To achieve this we will need a partition function, which uses a Hamiltonian to describe the statistical properties of a system, giving us crucial information about the system's dynamics. Using $\hat H_{\text{MF}}$ with the canonical partition function yields
\[
Z = e^{-\beta \hat H_{\text{MF}}},
\]
where $\beta$ is a particular thermodynamical quantity, $(k_B T)^{-1}$. Expanding the site-wise partition function and using some trigonometric identities, we have
\[
Z = e^{-\beta \frac{NzJm^2}{2}} 2\text{cosh}(\beta zJm)^N.
\]
Finally, to find our magnetisation $m$, we must minimise the free energy of the system with respect to $m$. Using $F=-(\beta N)^{-1}\ln\{Z\}$,
\[
F = \frac{1}{2}zJm^2 -\frac{1}{\beta} \ln\{ 2\text{cosh}(\beta zJm) \}.
\]
The $m$ which minimises this free energy, by solving $\pdv{F}{m}=0$, is
\[
m = \tanh\left( \frac{zJm}{k_B T} \right),
\]
where we have used the previous definition of the thermodynamic beta. If we define the critical temperature as $T_c = \frac{zJ}{k_B}$, then this simplifies to
\begin{equation}\label{MFmodel}
m  = \tanh\left( \frac{ T_c m}{T} \right),
\end{equation}
the plot of which is contained in Figure \ref{fig:m(T)}.

Immediately we observe a hyperbolic tangent function arise in this mean field model. The curve bifurcates at the critical value of $T$, showing the two possible magnetised states. We clearly see either $m = -1$ or $m = 1$, given by the ends of \eqref{MFmodel}. We disregard the zero solution at $T<T_c$ as energetically unfavourable.

Note that this $\tanh$ curve is a different sort of activation function; rather than determining magnetisation, it determines either of the possible magnetised states. In principle, we could maintain this bifurcation: suppose we defined two different firing patterns, where, upon receiving an input and crossing a critical point, all channels either contained an $\text{Na}^+$ ion or all channels did not. This could represent the firing of an inhibitory neurone causing selective inactivity in a firing excitatory neurone, which would normally communicate a signal. Then, we would have something corresponding to a critical input creating a single spike according to the statistics of the input. In this case, the activation function determines whether a stimulus is likely to elicit excitatory or inhibitory neural spikes, perhaps comprising a different sort of classification.

However, much like this is not main feature of the Ising model phase transition, this is not the major point of this paper. Instead, we restrict the magnetisation to $m = 1$, and examine the resultant analogy to firing dynamics. This will also allow us to determine the salient features of the previously mentioned class of activation functions.

\subsection{Recovering the Activation Function from the Ising Model}

\begin{figure}[htb!]
\resizebox{0.45\textwidth}{!}{
\includegraphics{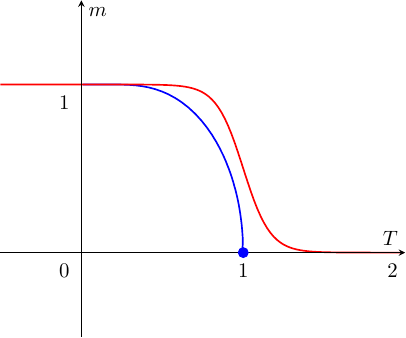}
}
\caption{\textbf{Magnetisation fit to a hyperbolic tangent.} The previously defined magnetisation curve can be fit to another hyperbolic tangent curve, satisfying the typical firing or not-firing activation function with $V$ increasing like $-T$. The hyperparameter $a$ is set to $a=6$, and $T_c = 1$. A second hyperparameter multiplying the critical temperature can be used to bring the fit even closer.}
\label{fig:fit}
\end{figure}

It is clear to see from Figure \ref{fig:m(T)} that, at temperatures above $T_c$, the only solution is $m=0$. Below $T_c,$ the solution is $m=-1$ or $1$, given by the two ends of \eqref{MFmodel}. When restricted to $m=1$, this curve behaves like a different hyperbolic tangent function, going from zero to one. So, for some parameter $a$, our function looks like
\begin{equation}\label{mag}
m(T) = -\frac{1}{2}\tanh\left(a(T-T_c)\right)+\frac{1}{2}
\end{equation}
which reverses when we set the temperature to neural state, as suggested earlier: recall the neural membrane voltage is itself negative, and so $T=-V$. As seen in Figure \ref{fig:fit}, this reproduces the neural activation function, where the threshold $T_c$ is a bias and the switching behaviour represents spiking or not spiking.

Since $T$, the temperature, depends on the quench, we can combine the temperature of the bath $T$ with the quench, $T(t) = T - \delta(t-t_c) Q.$
Here, $t_c$ is a time when cooling is applied, and the delta function $\delta(x)$ returns one for an argument of zero, and zero everywhere else. Thus, cooling only acts on the temperature when $t=t_c$. Note that, whilst an interaction with a thermal field could have been included in the Hamiltonian, we opt to couple it to the temperature later in the paper, for a variety of reasons\textemdash not in the least to follow through on our approximation of anomalously introduced local effects, and our desire to begin with a time-independent, and thus equilibrium, Hamiltonian. 

We may now parameterise motion along this curve due to changes in temperature in time. Recall we have negated temperature, as it is equal to neural membrane voltage, $-V$. We have already coupled our quench to temperature as a subtractive element that restores it to order. Suppose the model heats up linearly. This is accurate with respect to the neurone, which uses pumps to eject positive ions at a constant rate. Indeed, we have previously defined a quench as a perturbation from equilibrium. In reality, it is the influx of positive ions due to some firing event adjacent to the neural cell. The model will heat up again as soon as it loses heat, or pumps ions out of its cell body. We assume the neural pump acts with a constant speed, so that the time spent in the $m=1$ regime can be parameterised as linear in time $t$. Suppose also that this rate is one degree per second, such that $\dd{T} / \dd{t} = 1$ and $T(t) = t - Q$ for $t_c < t \leq Q$. Then, the number of spikes emitted, or the time spent in $m=1$, is the integral of \eqref{mag} from zero to $Q$. This is because, under our assumptions, the time to cool is equal to one kelvin per unit time. In that case, the time to cool back to zero perturbation when $Q$ is subtracted from $T$ is exactly $Q$.

Given the analogy drawn in Sections \ref{TwoA} and \ref{TwoB} holds, clearly, the mean field description of neural firing leads to a function that describes neural firing based on perturbations to equilibrium, as described above. Then, the number of spikes emitted in this time is the time spent in $m=1$, or the time before cooling, which is the previously described integral of \eqref{mag} with respect to temperature. This integral is evaluated as follows:
\begin{align*}
\frac{1}{2} \int_0^{Q} \left[ \text{tanh}(T)+1 \right] &\dd{T} = \phantom{}
\\&Q + \frac{1}{2}\ln\{2\} + \frac{1}{2}\ln\left\{1+e^{-2Q}\right\} + C.
\end{align*}

Clearly, we have the linear term dominating for $Q\gg0$; in which case, for $C=0$, we recover our ReLU function. This behaviour also reproduces that of the exponential linear unit for $C=-\alpha$, differing by no more than $\alpha$ anywhere. In general, we have an expression for linear or approximately linear, unbounded-above activation functions. 

If we choose to fit a sigmoid function to our magnetisation, rather than the hyperbolic tangent, then we have a similar result:
\[
\int_0^Q \frac{1}{1+e^{-T}}\dd{T} = Q+\ln\left\{1+e^{-Q}\right\} + C.
\]

The relevance of this with respect to neural firing is that the saturating activation function is only a binary classification case, with one spike\textemdash a single logic gate. More complex learning, such as the encoding of complex stimuli, on the other hand, requires many spikes. Hard quenches, or strong inputs, mean more time spent in the $m = 1$ regime; thus, stronger inputs mean more spikes get emitted. We then recover ReLU and ELU as functions for firing rate, by counting spikes over time. Since time spent magnetised, or time before heating, corresponds directly to quench strength, so too does spike count.

\section{Discussion}

To summarise our argument, we have shown that, for underlying physical dynamics which are complicated enough to perform inference at some (coarse-grained) level, the sigmoidal non-linearities of an activation function arise by simple scaling arguments from mean field theory. Conversely, scaling arguments reveal that sigmoidal activation functions are necessary for a model of neural dynamics which is capable of inference. This suggests a principled reason for the ubiquity of these two classes of activation function: artificial neural networks are physically realistic coarse-grainings of biological neural networks if and only if their firing dynamics look like sigmoid functions (or, when extended through time, rectifiers).

Given their comparatively small number of units and reduced computational power, one does indeed expect artificial neural networks to be mean fields of biological neural networks. In fact, it has been shown that one biological neurone can be modelled using multiple deep or artificial neural networks \cite{beniaguev2021single}, suggesting that many interacting mean field units are needed to recover the microscopic activity of a real neurone. We recall, for instance, \cite{bialek}, where a thermodynamic limit is postulated when making sense of the firing statistics of large neural networks (suggesting finite size effects in the absence of a large number of neurones); also \cite{bialek2}, where it is shown that pairwise correlations contain important information in biological neurones. That the design of traditional artificial neural networks follows from mean field analysis, and the consistency of this observation with previous results in the literature, confirms the general idea that biological neural networks are orders of magnitude more complex than their artificial counterparts. It remains to be seen how the analysis here lends itself to numerical estimates of the complexity of a biological neural network.

We now dissect some consequences of this argument.

% https://neurosciencenews.com/single-neuron-deep-learning-19264/

% https://www.tu.berlin/en/about/profile/press-releases-news/2022/februar/artificial-brain-with-a-single-neuron

%https://www.nature.com/articles/s41467-021-25427-4

%https://www.quantamagazine.org/how-computationally-complex-is-a-single-neuron-20210902/

\subsection{Firing Rates and Sparse Neural Codes}

Evidence suggests that neurones rely on sparse coding to efficiently communicate stimuli, especially in high-noise or high-dimensional environments. In fact, many separate neural coding schemes have been considered to emerge from sparsity, which neural networks employ due to energy constraints and to cope with dimensionality \cite{SparsePLOS}. Broadly, sparse coding states that different firing rates, which contain representations of information by encoding features of a stimulus, will be sparsely distributed in a neural network. In large neural populations, key neurones will be firing at various rates and most other neurones will not be firing at all. Such a sparse code is advantageous for efficient learning by decorrelating inputs, which allows features to be coded independently. Crucially, this leads to a robust representation, and is equivalent to reducing the coding of redundant features whilst preserving coded information \cite{SparseCyb}.

The rectifier, or the unbounded-above activation function we discuss, has indeed been shown to improve representation in deep neural networks by precisely these mechanisms \cite{SparseDNNs}. Some neurones are firing with a particular rate, lying on the linear portion of the curve, and others are resting, lying on the portion of the curve valued at zero. The coding benefits highlighted are exactly those found in sparse representations in biological neural networks, where disentangling is referred to as decorrelating inputs, which assists in learning high-dimensional data. These networks also utilise sparsity as a rich but energy efficient coding scheme, showing that in deep neural networks, sparse representations take fewer computational resources whilst showing high training accuracy. 

In our model, this sparsity is reproduced by local effects such as quenches of different magnitudes acting on particular neurones in the network. %for a distro of quench values, do we get wilson cowan back? 
It can be shown that, for Gaussian distributed quenches, our results imply the results in \cite{WilsonCowan1972}: calculating the transfer function of a network of neurones under quenches fluctuating about a mean of zero follows the argument in \cite{WilsonCowan1972} exactly.

\subsection{Energy-Based Learning}
Recently, a broad theory of machine learning and inference has been formulated in an energy-based framework\textemdash in particular, a paradigm based on energy minimisation has been proposed in \cite{EBL2006}, where choosing a network configuration that minimises energy is equivalent to finding an output that minimises loss. These follow on older ideas where free energy minimisation is employed in statistical learning, such as the Boltzmann machine or spin glass models. Here, the energy of a configuration is used as a penalty, following the idea that physical systems seek to minimise free energy and that this underlies the stability of a given state. This appeals to statistical mechanical ideas about energy minimisation, which we have already used in discussing the Ising model\textemdash the configuration chosen by a system always obeys a minimisation principle. As such, this can be used as a measurement of error, where we designate high energy states as being incorrect in both the physical and statistical sense.

A useful way of thinking about the idea of free energy minimisation is that free energy is defined as
\[
F = E - TS.
\]
For clamped energy levels, clearly, maximising entropy is equivalent to minimising free energy, since 
\[\Delta F = -T \Delta S\]
for constant energy and temperature. Then, free energy minimisation is a natural consequence of the second law of thermodynamics, which states that systems will always produce greater entropy. In the information theoretic sense, defined in \cite{JaynesMaxEnt} as essentially equivalent to the thermodynamical sense, maximising entropy is choosing the most best model of observed variables. Thus, we have a direct application to our inference or learning process. 

Following this, we examine why a neural Ising model spikes. Clearly, when the temperature decreases, the entropic contribution to free energy decreases as well. Hence, minimisation of free energy occurs when total energy is minimised. We observed this happen when the Hamiltonian was in a magnetised state. In the sense of an error signal, when an input\textemdash a temperature lowering quench\textemdash arrives, the error in the system is high as long as the Ising model occupies a high energy state, which is unlikely given the physical and statistical scenario. By magnetising, or spiking, the system decreases this error through responding to the input, which is equivalent to choosing a free energy minimising stable state. In the energy-based learning scheme, loss functions are often arrived at by explicitly considering the marginalised Gibbs distribution over the inputs to the system, and learning is performed by minimising the resultant free energy in the zero-temperature limit.

This accords with other, more biological ideas concerning energy minimisation in learning, wherein neural spikes learn the relationship between stimulus and evoked response, and minimisation of energy underlies learning. It has been found that real neural networks, {\it in vitro}, minimise \emph{variational} free energy when learning representations of stimuli \cite{FEPin-vitro}. Variational free energy is an information-theoretic notion closely related to the thermodynamical Helmholtz free energy, although whether only by statistical mechanical analogy or also by physical principles remains controversial \cite{KieferFEP,Andrews2020}.

\subsection{Self-Similarity and Criticality}

We note one final implication by suggesting a relationship between this result, and mean field theory applied to neural populations. In particular, we note that to recover non-linear firing statistics, the collective dynamics of neural populations are almost ubiquitously described using a sigmoid function \cite{CorticalFields}. The importance of the sigmoid function in statistical approximations of neural population dynamics\textemdash especially mean field models\textemdash was first suggested in \cite{WilsonCowan1972}, wherein it was shown that for a realistic model of population firing, the proportion of firing cells naturally followed a sigmoid. In following work by Amari, non-linear functions were also necessary to model the collective dynamics of a neural field as a self-organised pattern \cite{amari}. More recently, transfer functions in biologically-realistic mean field models have taken the form of a rectifier \cite{destexhe}, corresponding to our own unification of unbounded-above functions with sigmoid-type functions as a firing rate. The results in this paper undoubtedly extrapolate to the case of neurones as a subunit and neural populations as a mean field, a relationship consistent with the approximate self-similarity observed in the human cortex. We note that self-similar systems, including the Ising model around $T_c$, are generally in a state of criticality; this is the so-called `edge of chaos' close to a phase transition. Signatures of criticality have been observed in the brain \cite{BrainCrit,BreakspearCrit}, and critical dynamics are known to be important for computation in both biological and artificial neural networks, which have been shown to perform best at criticality \cite{pmlr, bialek, EdgeOfChaos,neural-net-crit}. This congruence adds a dimension to these results, as they capture the dependence of ideal computation on the scale-invariance of certain expressions.

\section{Conclusion}
We have shown that, as a model of the key features of real neural dynamics, an artificial neural network is a mean field model of biological neural networks. This model falls in the Ising universality class, and thus an artificial neural network naturally exhibits a sigmoidal or $\tanh$-like switching behaviour between firing and not firing. Various conventional activation functions can be easily arrived at from this mean field model; as such, we have motivated the designs of historical and modern artificial neural networks, and in particular, the concept and typical form of the activation function. In so doing, we have also examined how ideal learning necessarily invokes the non-linear processes in the neurone, and utilises energy minimisation, by modelling this process with an Ising model and applying other statistical mechanical ideas.

\bibliographystyle{unsrt}
\bibliography{main}

\begin{thebibliography}{10}

\bibitem{Abraham}
Tara~H Abraham.
\newblock ({P}hysio)logical circuits: the intellectual origins of the
  {M}c{C}ulloch–{P}itts neural network.
\newblock {\em Journal of the History of the Behavioral Sciences}, 38(1):3--25,
  2002.

\bibitem{MP1943}
Walter McCulloch and Warren Pitts.
\newblock A logical calculus of the ideas immanent in nervous activity.
\newblock {\em Bulletin of Mathematical Biophysics}, 5:115--133, 1943.

\bibitem{WilsonCowan1972}
Hugh~R Wilson and Jack~D Cowan.
\newblock Excitatory and inhibitory interactions in localized populations of
  model neurons.
\newblock {\em Biophysical journal}, 12(1):1--24, 1972.

\bibitem{NeurIPS2012}
Alex Krizhevsky, Ilya Sutskever, and Geoffrey~E Hinton.
\newblock Imagenet classification with deep convolutional neural networks.
\newblock In F~Pereira, C~J~C Burges, L~Bottou, and K~Q Weinberger, editors,
  {\em Advances in Neural Information Processing Systems}, volume~25, pages
  1097--1105. Curran Associates, Inc., 2012.

\bibitem{ICML}
Vinod Nair and Geoffrey~E Hinton.
\newblock Rectified linear units improve restricted {B}oltzmann machines.
\newblock In {\em Proceedings of the 27th International Conference on Machine
  Learning}, ICML'10, page 807–814, Madison, WI, USA, 2010. Omnipress.

\bibitem{pmlr}
Soufiane Hayou, Arnaud Doucet, and Judith Rousseau.
\newblock On the impact of the activation function on deep neural networks
  training.
\newblock In Kamalika Chaudhuri and Ruslan Salakhutdinov, editors, {\em
  Proceedings of the 36th International Conference on Machine Learning},
  volume~97 of {\em Proceedings of Machine Learning Research}, pages
  2672--2680. PMLR, 2019.

\bibitem{pinkus1993}
Moshe Leshno, Vladimir~Y Lin, Allan Pinkus, and Shimon Schocken.
\newblock Multilayer feedforward networks with a nonpolynomial activation
  function can approximate any function.
\newblock {\em Neural Networks}, 6:861--867, 1993.

\bibitem{Ising67}
Stephen Brush.
\newblock History of the {L}enz-{I}sing model.
\newblock {\em Reviews of Modern Physics}, 39(4):883--893, 1967.

\bibitem{liq-gas-crit}
Alistair~D Bruce and Nigel~B Wilding.
\newblock Scaling fields and universality of the liquid-gas critical point.
\newblock {\em Physical Review Letters}, 68(2):193--196, Jan 1992.

\bibitem{IsingStrings}
Ara Sedrakyan.
\newblock 3{D} {I}sing model as a string theory in three-dimensional
  {E}uclidean space.
\newblock {\em Physics Letters B}, 304(3):256 -- 262, 1993.

\bibitem{HH1952}
Alan~L Hodgkin and Andrew~F Huxley.
\newblock A quantitative description of membrane current and its application to
  conduction and excitation in nerve.
\newblock {\em The Journal of Physiology}, 117(4):500--544, 1952.

\bibitem{beniaguev2021single}
David Beniaguev, Idan Segev, and Michael London.
\newblock Single cortical neurons as deep artificial neural networks.
\newblock {\em Neuron}, 109(17):2727--2739, 2021.

\bibitem{bialek}
Gašper Tkačik, Thierry Mora, Olivier Marre, Dario Amodei, Stephanie~E Palmer,
  Michael~J Berry, and William Bialek.
\newblock Thermodynamics and signatures of criticality in a network of neurons.
\newblock {\em Proceedings of the National Academy of Sciences},
  112(37):11508--11513, 2015.

\bibitem{bialek2}
Elad Schneidman, Michael~J Berry, Ronen Segev, and William Bialek.
\newblock Weak pairwise correlations imply strongly correlated network states
  in a neural population.
\newblock {\em Nature}, 440(7087):1007--1012, 2006.

\bibitem{SparsePLOS}
Michael Beyeler, Emily~L Rounds, Kristofor~D Carlson, Nikil Dutt, and Jeffrey~L
  Krichmar.
\newblock Neural correlates of sparse coding and dimensionality reduction.
\newblock {\em PLOS Computational Biology}, 15(6):1--33, 06 2019.

\bibitem{SparseCyb}
Peter F\"old\'iak.
\newblock Forming sparse representations by local anti-{H}ebbian learning.
\newblock {\em Biological Cybernetics}, 64:165--170, 1990.

\bibitem{SparseDNNs}
Xavier Glorot, Antoine Bordes, and Yoshua Bengio.
\newblock Deep sparse rectifier neural networks.
\newblock In Geoffrey Gordon, David Dunson, and Miroslav Dudík, editors, {\em
  Proceedings of the Fourteenth International Conference on Artificial
  Intelligence and Statistics}, volume~15 of {\em Proceedings of Machine
  Learning Research}, pages 315--323. JMLR Workshop and Conference Proceedings,
  2011.

\bibitem{EBL2006}
Yann Le{C}un, Sumit Chopra, Raia Hadsell, {Marc Aurelio} Ranzato, and {Fu Jie}
  Huang.
\newblock A tutorial on energy-based learning.
\newblock In G~Bakir, T~Hofman, B~Scholkopt, A~Smola, and B~Taskar, editors,
  {\em Predicting structured data}. MIT Press, 2006.

\bibitem{JaynesMaxEnt}
Edwin~T Jaynes.
\newblock Information theory and statistical mechanics.
\newblock {\em Physical Review}, 106(4):620--630, 1957.

\bibitem{FEPin-vitro}
Takuya Isomura and Karl Friston.
\newblock In vitro neural networks minimise variational free energy.
\newblock {\em Nature Scientific Reports}, 8(1):1--14, 2018.

\bibitem{KieferFEP}
Alex~B Kiefer.
\newblock Psychophysical identity and free energy.
\newblock {\em Journal of the Royal Society Interface}, 17:20200370, 2020.

\bibitem{Andrews2020}
Mel Andrews.
\newblock The math is not the territory: navigating the free energy principle.
\newblock {\em Biology \& Philosophy}, 36(30), 2021.

\bibitem{CorticalFields}
Gustavo Deco, Viktor~K Jirsa, Peter~A Robinson, Michael Breakspear, and Karl
  Friston.
\newblock The dynamic brain: from spiking neurons to neural masses and cortical
  fields.
\newblock {\em PLOS Computational Biology}, 4(8):1--35, 08 2008.

\bibitem{amari}
Shun{-}ichi Amari.
\newblock Dynamics of pattern formation in lateral-inhibition type neural
  fields.
\newblock {\em Biological Cybernetics}, 27(2):77--87, 1977.

\bibitem{destexhe}
Matteo di~Volo, Alberto Romagnoni, Cristiano Capone, and Alain Destexhe.
\newblock Biologically realistic mean-field models of conductance-based
  networks of spiking neurons with adaptation.
\newblock {\em Neural Computation}, 31(4):653--680, 2019.

\bibitem{BrainCrit}
Zhengyu Ma, Gina~G Turrigiano, Ralf Wessel, and Keith~B Hengen.
\newblock Cortical circuit dynamics are homeostatically tuned to criticality in
  vivo.
\newblock {\em Neuron}, 104(4), 2019.

\bibitem{BreakspearCrit}
Luca Cocchi, Leonardo~L Gollo, Andrew Zalesky, and Michael Breakspear.
\newblock Criticality in the brain: a synthesis of neurobiology, models and
  cognition.
\newblock {\em Progress in Neurobiology}, 158:132--152, 2017.

\bibitem{EdgeOfChaos}
Ge~Yang and Samuel Schoenholz.
\newblock Mean field residual networks: on the edge of chaos.
\newblock In {\em Advances in Neural Information Processing Systems},
  volume~30, pages 7103--7114. Curran Associates, Inc., 2017.

\bibitem{neural-net-crit}
Stefan Landmann, Lorenz Baumgarten, and Stefan Bornholdt.
\newblock Self-organized criticality in neural networks from activity-based
  rewiring.
\newblock {\em Physical Review E}, 103(3):032304, Mar 2021.

\end{thebibliography}

\end{document}